\documentclass[11pt]{article}
\usepackage{graphicx}
\usepackage{xspace}
\usepackage{afterpage}

\newcommand\pubnumber{}
\newcommand\pubdate{\today}

\def\institute{University of Rochester}

\def\Title#1{\begin{center} {\Large #1 } \end{center}}
\def\Author#1{\begin{center}{ \sc #1} \end{center}}
\def\Address#1{\begin{center}{ \it #1} \end{center}}

\newcommand\pubblock{\rightline{\begin{tabular}{l} \pubnumber\\
         \pubdate  \end{tabular}}}
\newenvironment{Abstract}{\begin{quotation}  }{\end{quotation}}
\newenvironment{Presented}{\begin{quotation} \begin{center} 
             PRESENTED AT\end{center}\bigskip 
      \begin{center}\begin{large}}{\end{large}\end{center} \end{quotation}}

\def\beq{\begin{equation}}
\def\eeq#1{\label{#1}\end{equation}}
\def\eeqn{\end{equation}}

\def\beqa{\begin{eqnarray}}
\def\eeqa#1{\label{#1}\end{eqnarray}}
\def\eeqan{\end{eqnarray}}

\let\bar=\overbar

\def\etal{{\it et al.}}

\def\Dslash{\not{\hbox{\kern-4pt $D$}}}
\def\dslash{\not{\hbox{\kern-2pt $\del$}}}

\def\msb{{\bar{\ssstyle M \kern -1pt S}}}

\newcommand{\ttbar}{\ensuremath{\mathrm{t\bar{t}}}\xspace}
\newcommand{\tq}{\ensuremath{\mathrm{t}\xspace}}

\newcommand{\pt}{\ensuremath{p_\mathrm{T}}\xspace}

\newcommand{\POWHEG}{\textsc{Powheg}\xspace}

\newcommand{\PYTHIA}{\textsc{Pythia}\xspace}
\newcommand{\HERWIG}{\textsc{HERWIG++}\xspace}

\newcommand{\lpj}{$\ell$+jets\xspace}
\newcommand{\DRtopjets}{\ensuremath{\Delta R_{\mathrm{j}_\tq}}\xspace}
\begin{document}
\begin{titlepage}
\pubblock

\vfill
\Title{Measurements of differential $t\bar{t}$ production cross sections at CMS}
\vfill
\Author{Otto Hindrichs\\For the CMS Collaboration}
\Address{\institute}
\vfill
\begin{Abstract}
An overview of recent measurements of differential top quark pair production cross sections performed by the CMS experiment at the LHC is presented. Measurements at different proton-proton center-of-mass energies are available using the dilepton, lepton+jets, and all-jets decay channels of the top quark. In addition to the measurements of parton-level top quarks, many measurements at particle level in an experimental accessible phase space are now available. For these results the dependence on theoretical extrapolations is reduced. A common observation of all measurements is a softer transverse momentum of the top quark than predicted by state of the art standard model calculations. However, new calculations with NNLO QCD and NLO electro-weak precision show an improved agreement.
\end{Abstract}
\vfill
\begin{Presented}
$10^{th}$ International Workshop on Top Quark Physics\\
Braga, Portugal,  September 17--22, 2017
\end{Presented}
\vfill
\end{titlepage}
\def\thefootnote{\fnsymbol{footnote}}
\setcounter{footnote}{0}

\section{Introduction}
The high production rate of top quark pairs (\ttbar) at the LHC allows for precise measurements of differential and double-differential cross sections as a function of kinematic variables of the top quark and the \ttbar system. Extensive measurements were performed by the CMS\,\cite{CMS} experiment at different proton-proton center-of-mass energies and in various \ttbar decay channels. These measurements provide an important test of the standard model and its ability to predict the observed distributions, which are sensitive to higher order perturbative QCD and electro-weak corrections~\cite{NNLOEW}. Moreover, the generation of \ttbar events requires a realistic modeling of the parton shower. Measurements of kinematic properties and multiplicities of jets associated with \ttbar production allow for a detailed comparison of different parton shower models to the data and provide insight into their tuning.

\section{Differential measurements at parton level}
For the parton level measurements the top quark is defined as a top quark directly before its decay, but after radiation. Since in this definition the decay products are not considered, the results are given without any kinematic restrictions. This requires a large extrapolation from quantities like jets and leptons measured in a restricted experimentally accessible phase space. Such an extrapolation relies on theoretical models that imply additional systematic uncertainties. However, a parton level measurement is the only way to compare the most precise fixed order theoretical predictions to the measurements.

\begin{figure}[htb]
\centering
\includegraphics[width=0.34\textwidth]{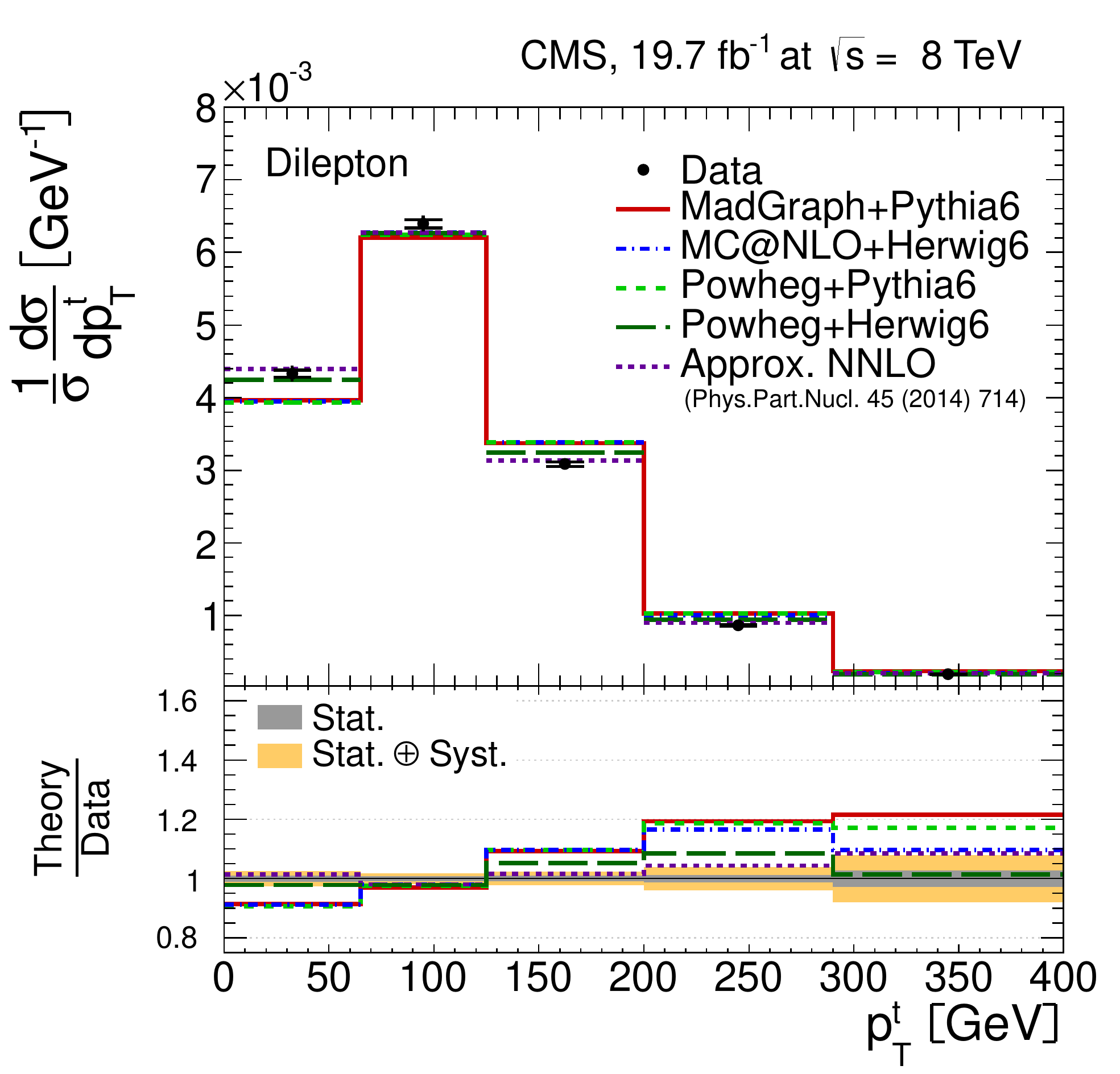}
\includegraphics[width=0.34\textwidth]{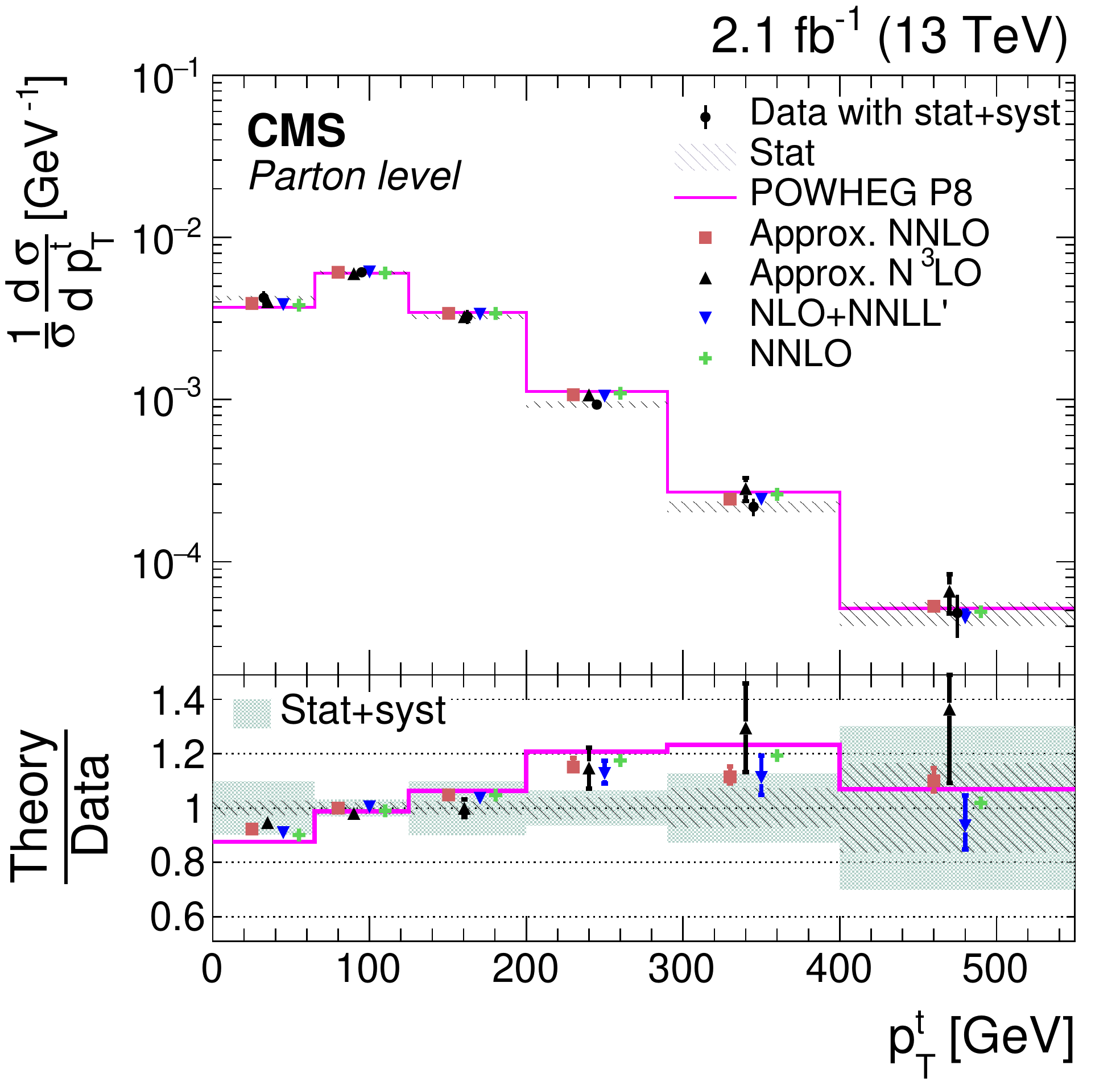}\\
\caption{Differential measurements as function of $\pt(\tq)$ at parton level in the dilepton channels at 8\,TeV~\cite{dilep8} and 13\,TeV~\cite{dilep13}.}
\label{fig_dilep}
\end{figure}

In Fig.~\ref{fig_dilep} results of the parton-level measurements of the differential cross section as a function of the transverse momentum of the top quark $\pt(\tq)$ are shown at center-of-mass energies of 8\,TeV~\cite{dilep8} and 13\,TeV~\cite{dilep13}. For these measurements in the dilepton decay channels, events with pairs of oppositely charged leptons ($e^+e^-$, $\mu^+\mu^-$, $e^\pm \mu^\mp$) and b jets are selected. After excluding a dilepton mass window around the Z boson mass in the same flavor channels a pure selection of \ttbar events is obtained. An algebraic algorithm based on constrains of top quark, W boson masses, and the event \pt balance is used to reconstruct the \ttbar system.

\begin{figure}[htb]
\centering
\includegraphics[width=0.32\textwidth]{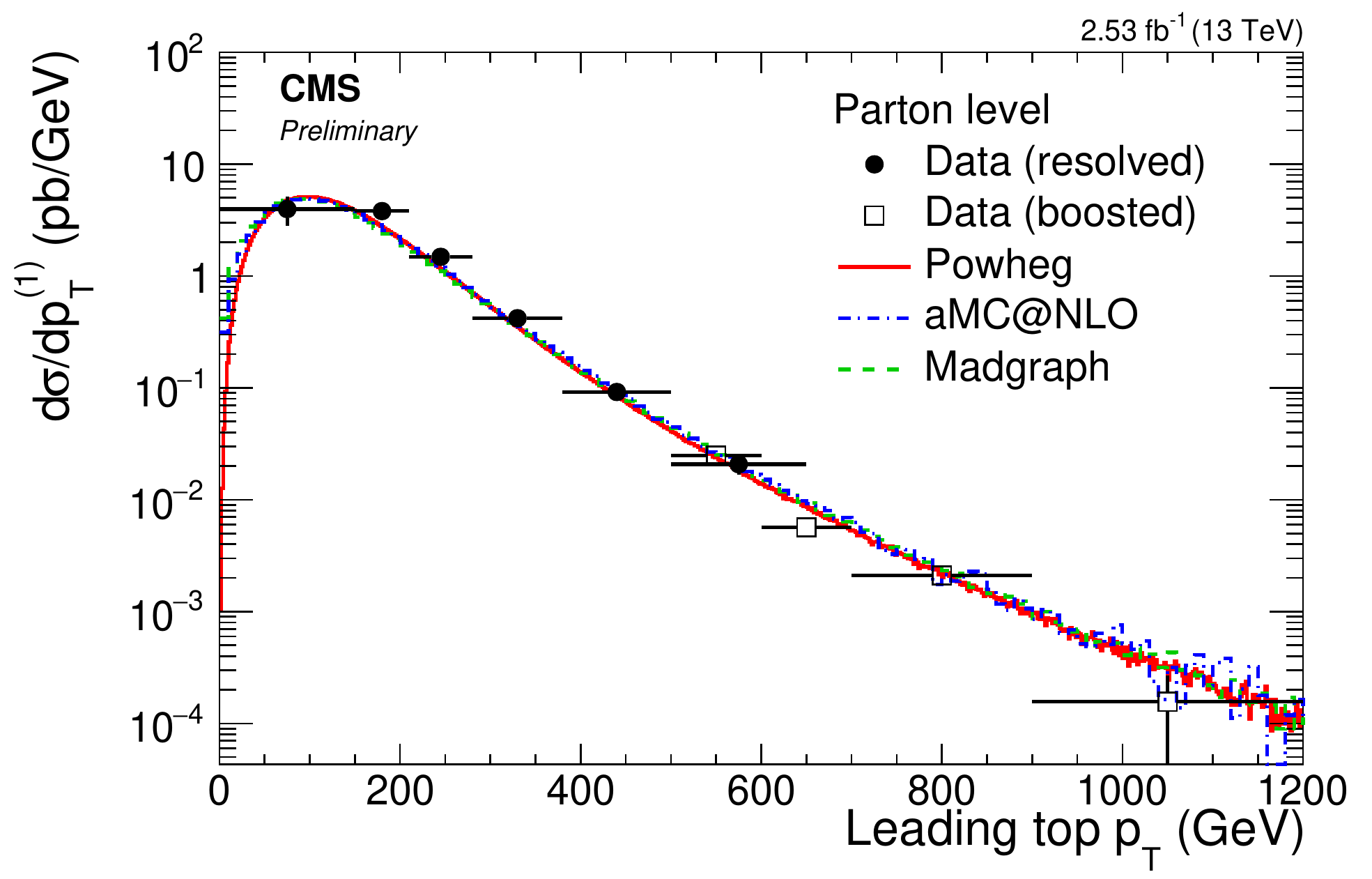}
\includegraphics[width=0.28\textwidth]{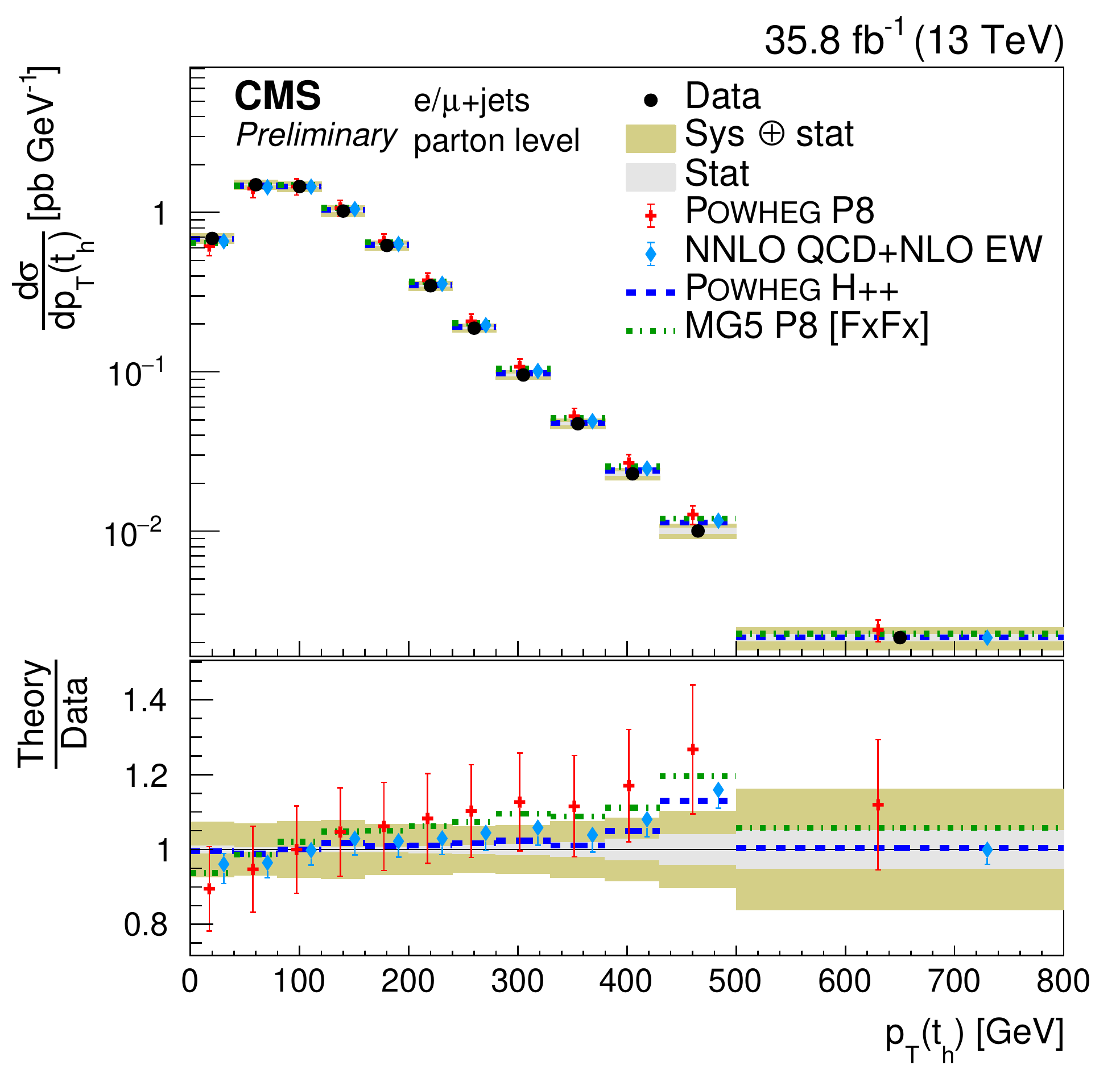}
\includegraphics[width=0.32\textwidth]{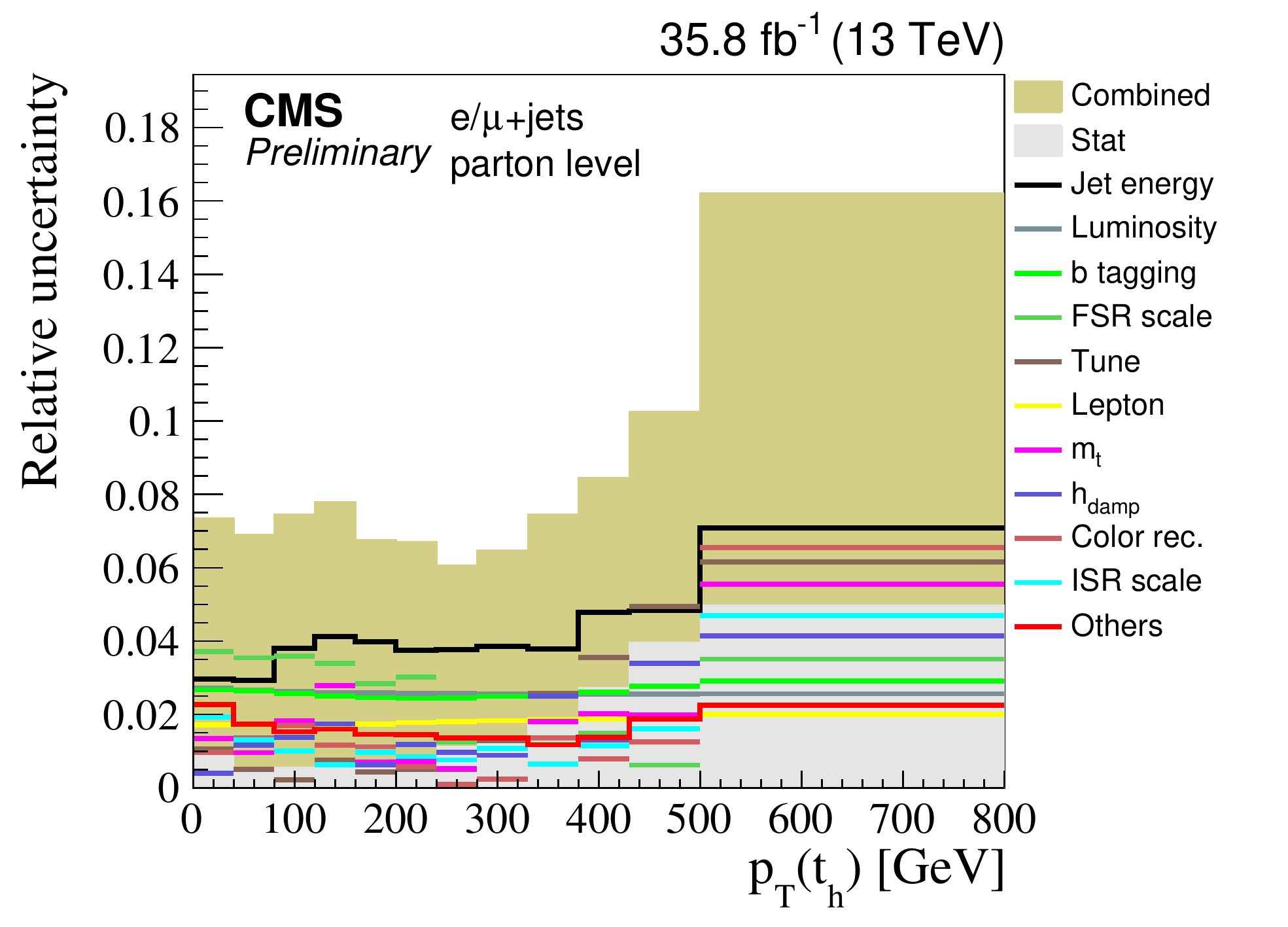}
\caption{Differential measurements as function of $\pt(\tq)$ at parton level in the all-jets~\cite{allhad13} (left) and \lpj~\cite{lepjet13} (middle) channels at 13\,TeV. The right panel shows a detailed decomposition of the uncertainty sources~\cite{lepjet13}.}
\label{fig_lepjet}
\end{figure}

In Fig.~\ref{fig_lepjet} results of a parton-level measurement in the all-jets and the \lpj channels at 13\,TeV are shown. For the measurement in the all-jets channel~\cite{allhad13} resolved and boosted top quarks are reconstructed. In the resolved case at least 6 jets are required, two must be identified as b jets. A kinematic fit is used to identify the top quark decay products. In the boosted case events containing two jets with $\pt > 250$\,GeV and 450\,GeV are selected. In each jet a b-tagged subjet must exist and their softdrop mass and n-subjettiness must be compatible with the decay of a boosted top quark. A template fit of the top quark mass distributions is used to extract the signal yield from the multijet background. For the measurement in the \lpj channel~\cite{lepjet13} events with a single electron or muon together with four jets are selected, where two jet must be identified as b jets. A likelihood based approach based on top quark and W boson mass constrains is used to reconstruct the \ttbar system. In all channels the measured $\pt(\tq)$ is softer than predicted by the calculations. However, taking into account the uncertainties on the experimental and theoretical side no contradiction to the standard model can be claimed. 

\begin{figure}[htb]
\centering
\includegraphics[width=0.31\textwidth]{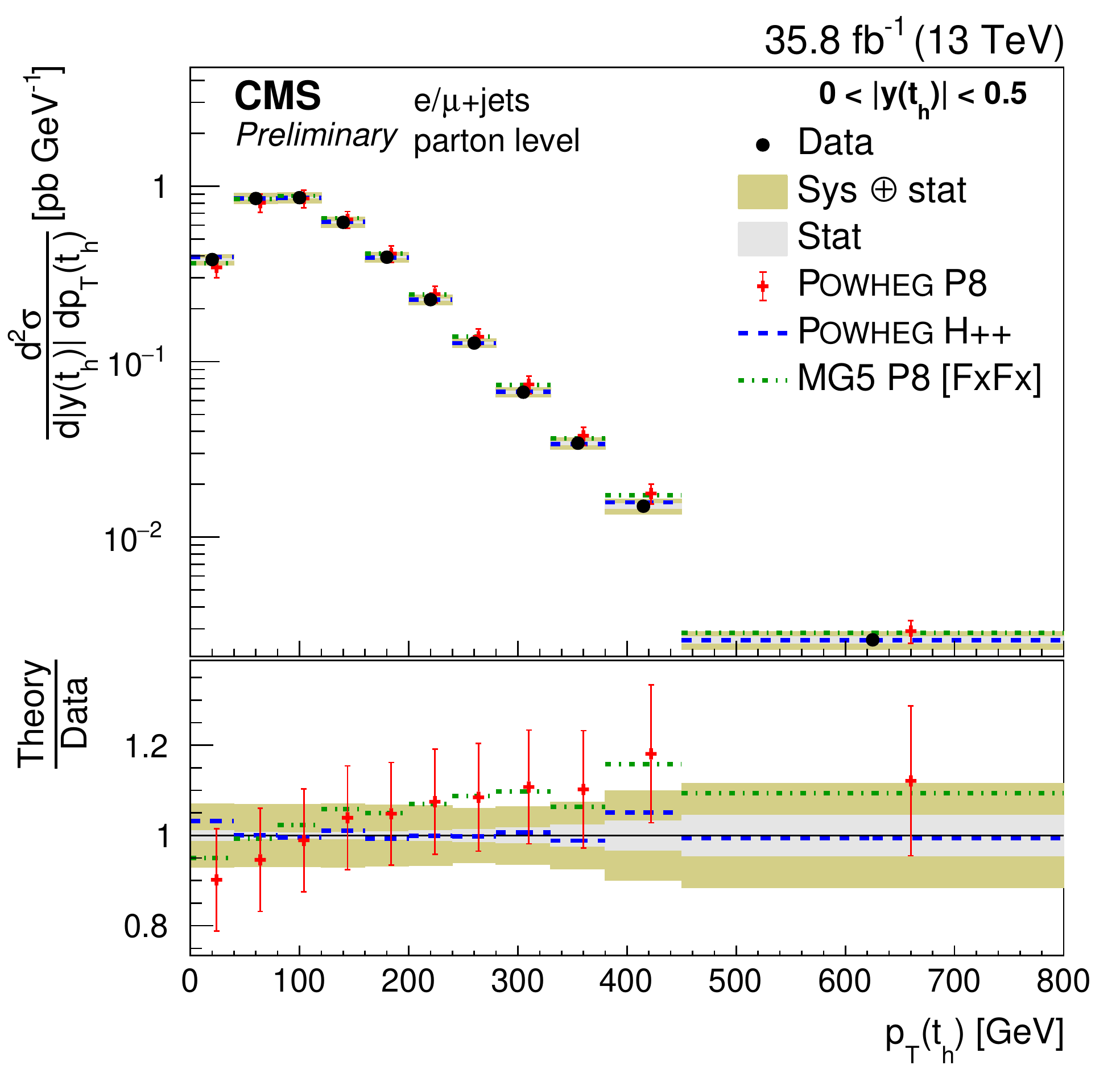}
\includegraphics[width=0.31\textwidth]{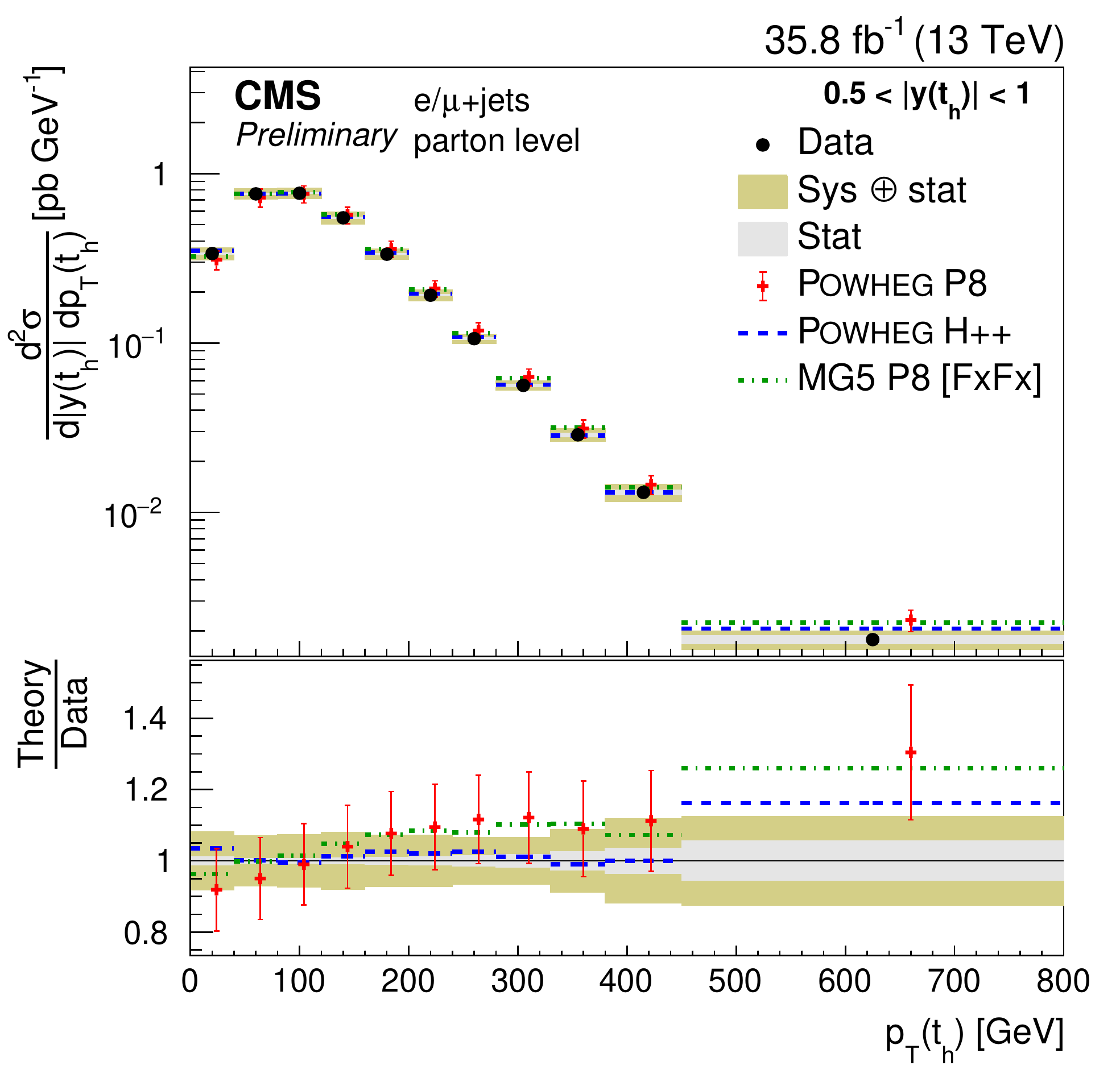}
\includegraphics[width=0.31\textwidth]{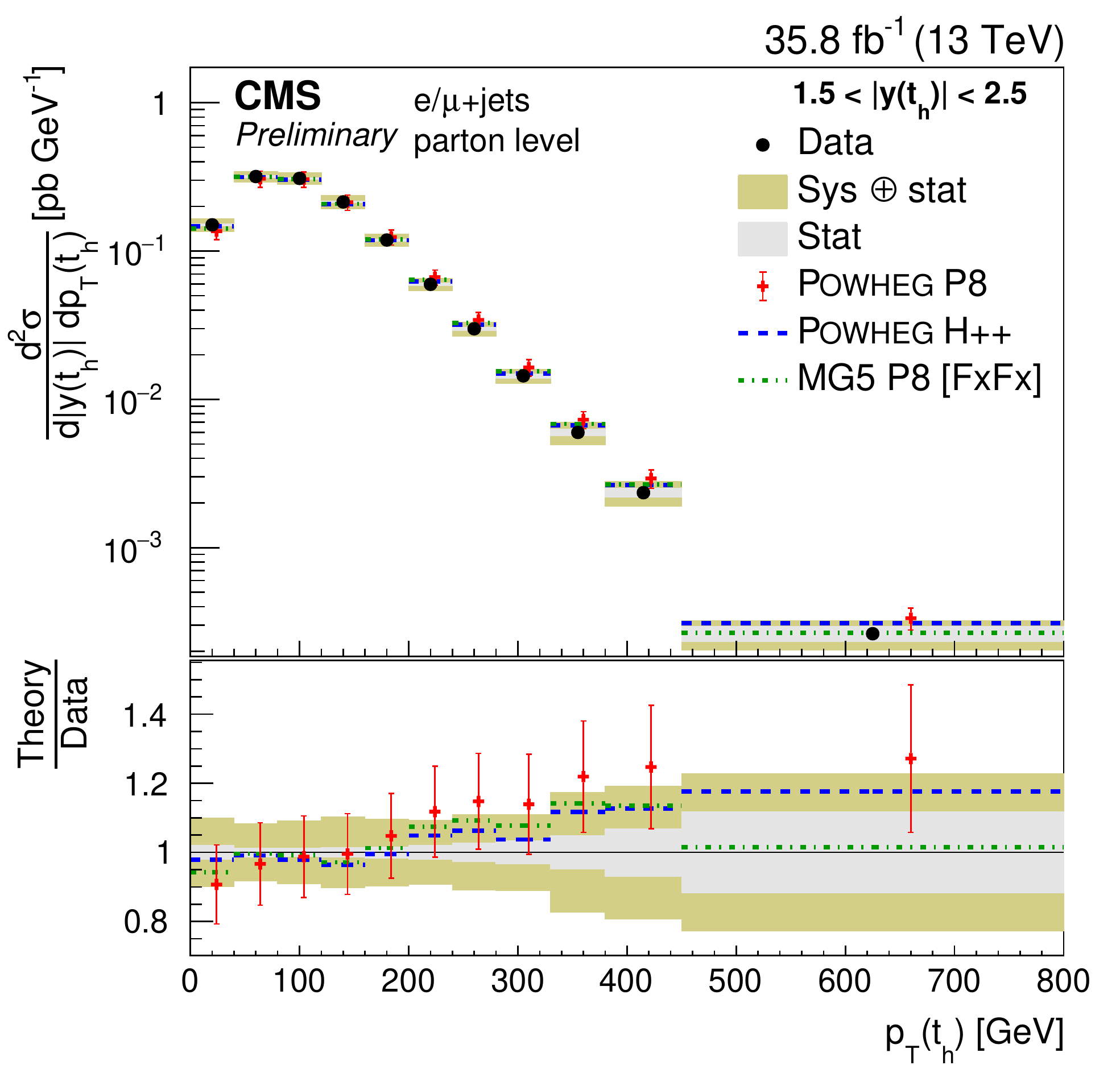}
\caption{Double-differential measurement as a function of $\pt(\tq)$ in regions of $y(\tq)$ at parton level in the \lpj channels at 13\,TeV~\cite{lepjet13}.}
\label{fig_doubledilep}
\end{figure}

Based on the same selection and reconstruction methods double-differential cross section in the \lpj channels are obtained. The result as function of $\pt(\tq)$ in regions of rapidity $y(\tq)$ is shown in Fig.~\ref{fig_doubledilep}. The measured softer $\pt(\tq)$ is persistent in all rapidity regions. The double-differential measurements in the dilepton channels at 8\,TeV~\cite{dilepdouble8} provide stronger constrains on the parton distribution function of the gluon than the one-dimensional measurements.

\section{Differential measurements at particle level}

\begin{figure}[htb]
\centering
\includegraphics[width=0.29\textwidth]{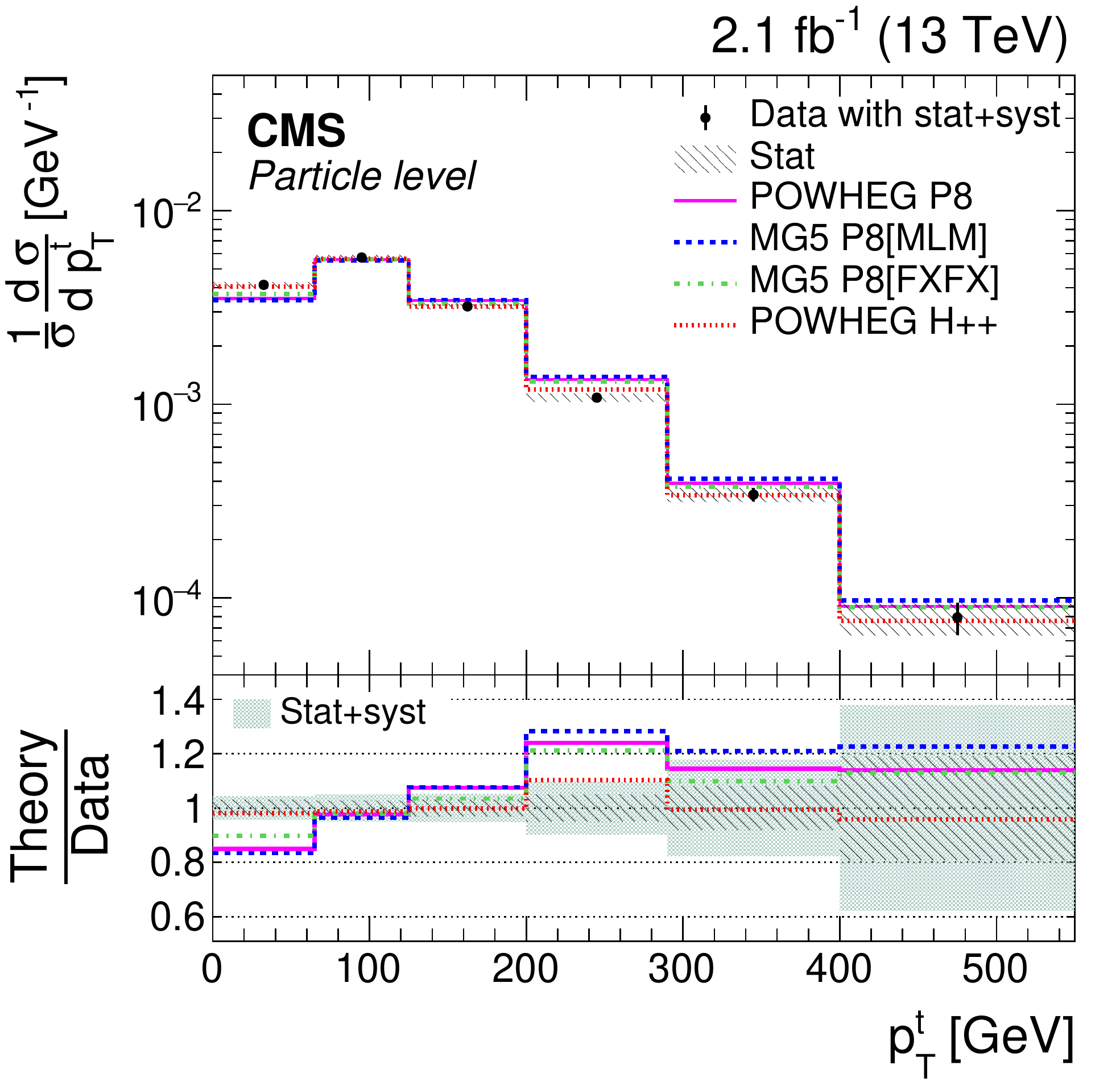}
\includegraphics[width=0.29\textwidth]{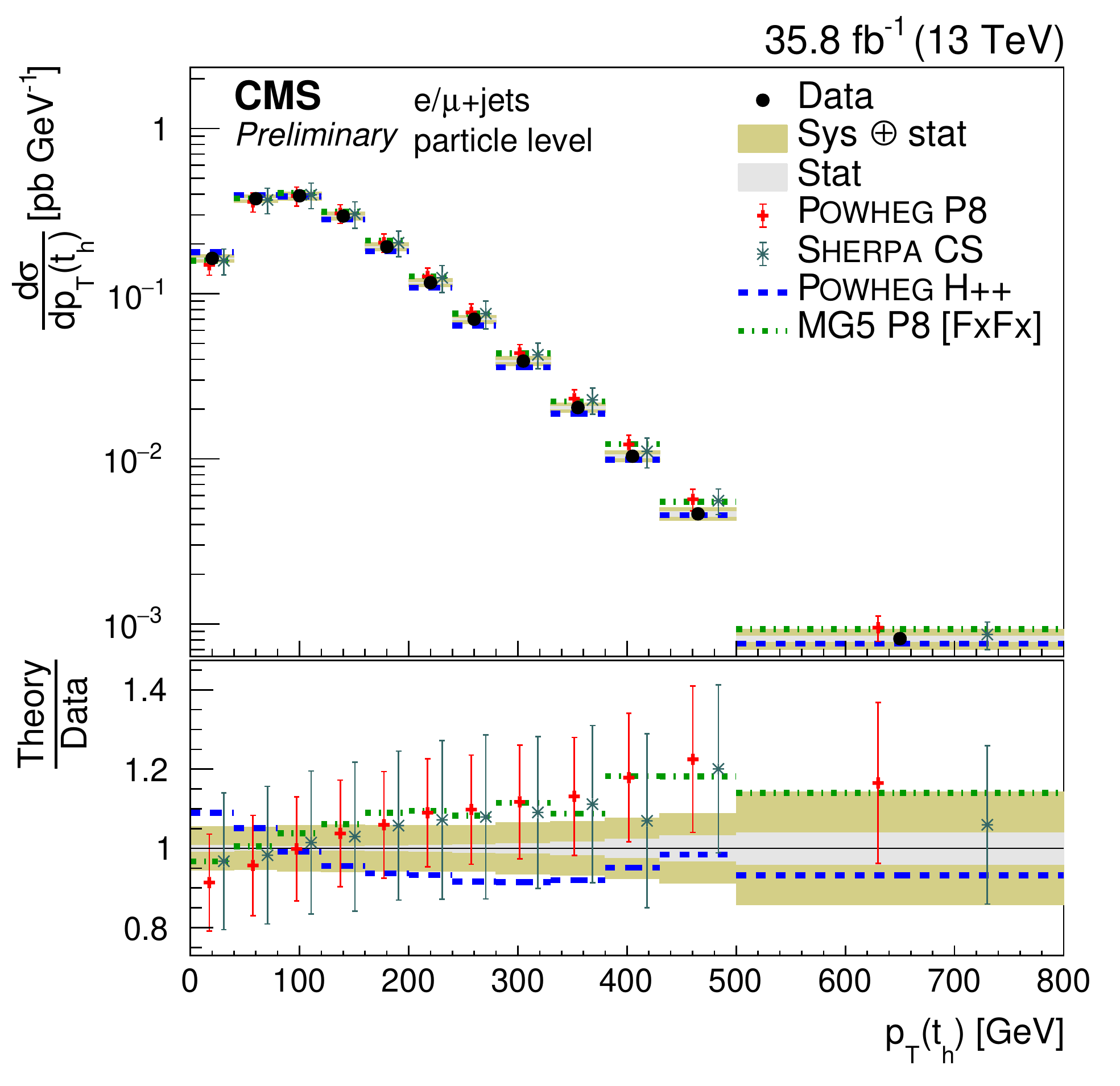}
\includegraphics[width=0.39\textwidth]{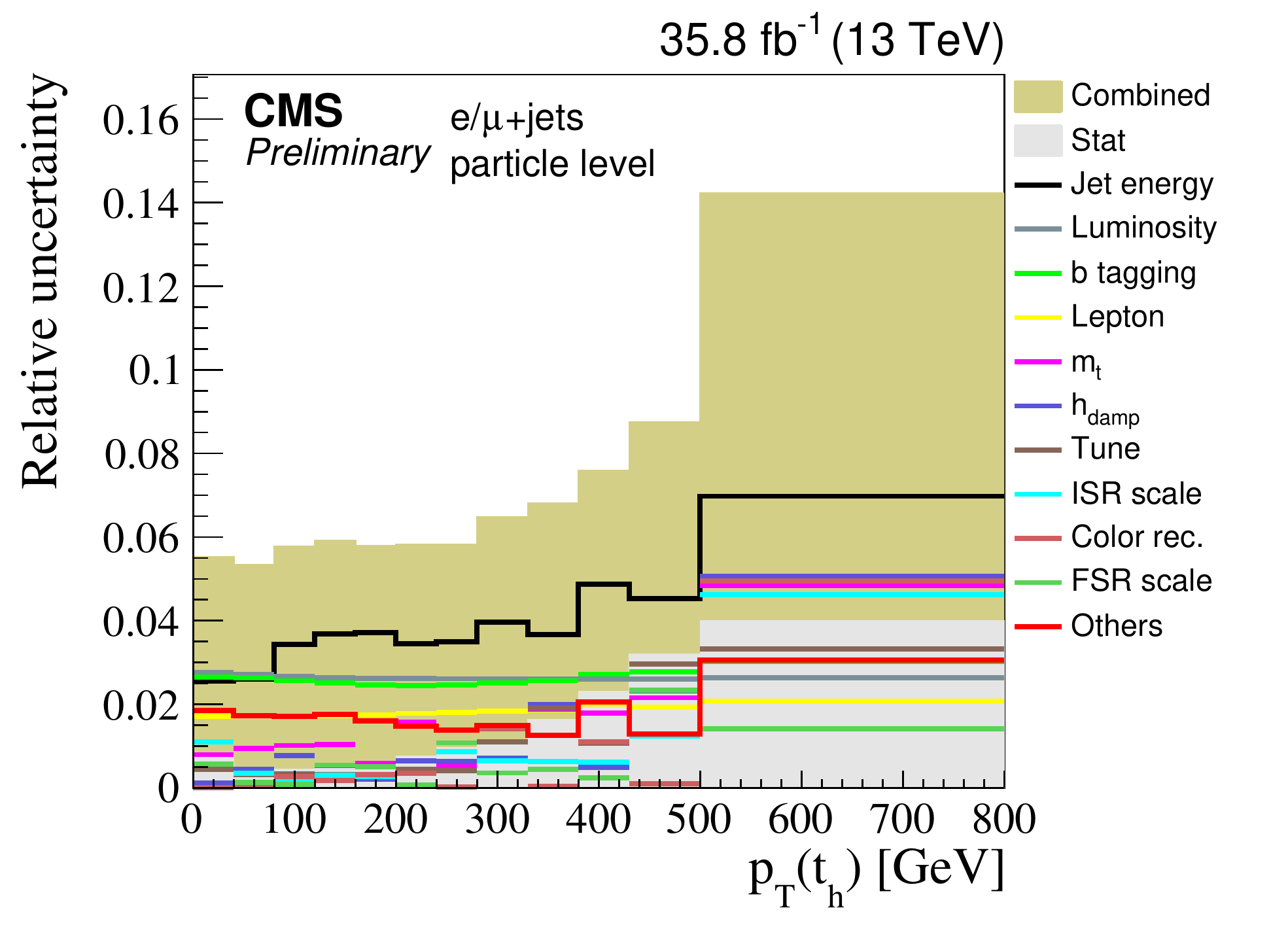}
\caption{Differential measurements as function of $\pt(\tq)$ at particle level in the dilepton channels (left)~\cite{dilep13} and in the \lpj channels (middle)~\cite{lepjet13} at particle level. The right panel shows a detailed decomposition of the uncertainty sources~\cite{lepjet13}.}
\label{fig_pla}
\end{figure}

For the measurements at particle level, proxies of the top quarks are defined based on long-living particles within the detector acceptance. Detailed definitions can be found in~\cite{dilep13, lepjet13, pl}. In Fig.~\ref{fig_pla} measurements as a function of $\pt(\tq)$ are shown in the dilepton and the \lpj channels. Again a softer $\pt(\tq)$ spectrum is observed. A comparison of the systematic uncertainties in the particle-level measurement and those in the parton-level measurement (Fig.~\ref{fig_lepjet}) shows a reduction of uncertainty. Especially, theoretical uncertainties like the effect of scale variations in the parton shower are reduced.   

\begin{figure}[htb]
\centering
\includegraphics[width=0.32\textwidth]{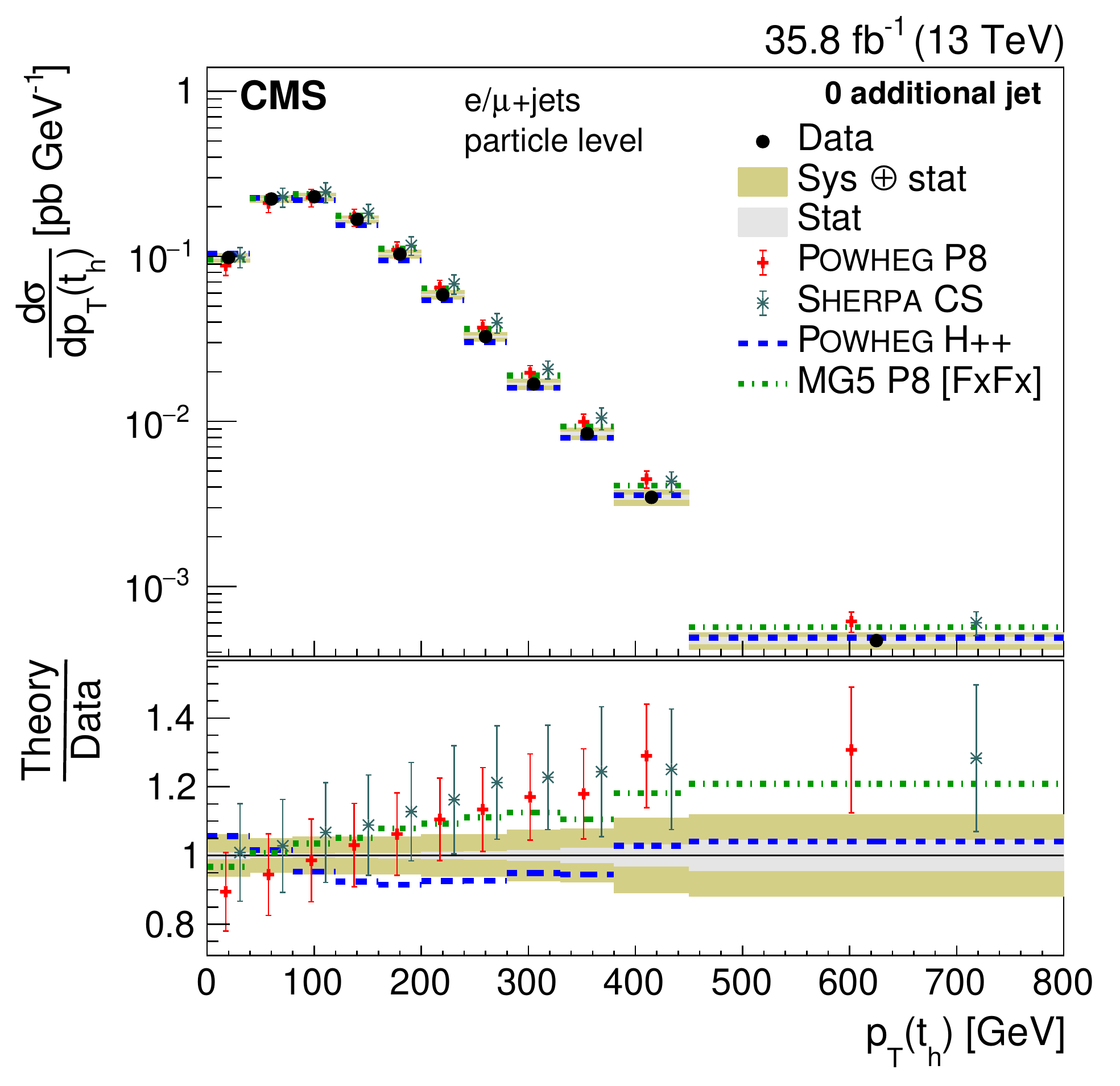}
\includegraphics[width=0.32\textwidth]{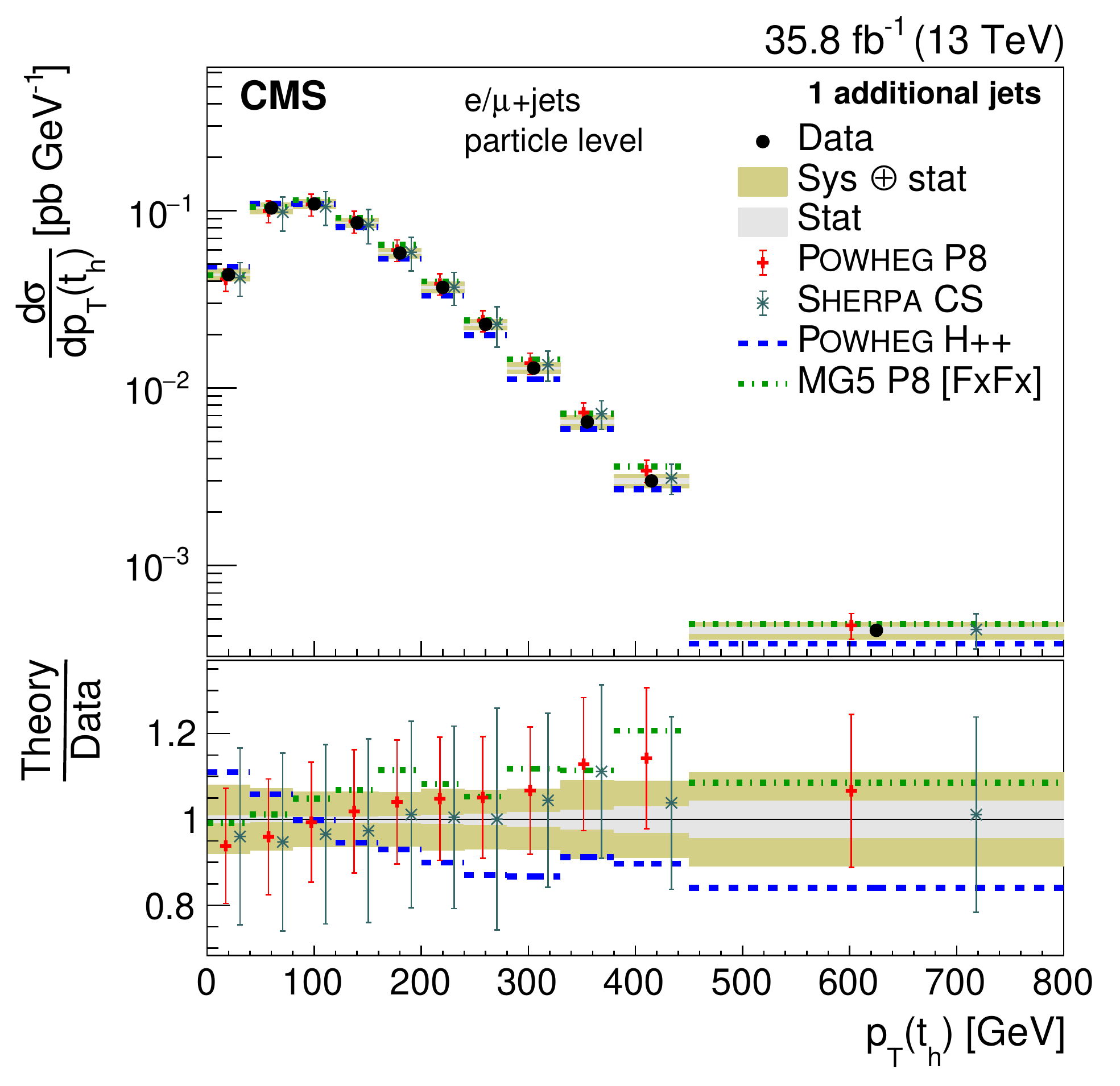}
\includegraphics[width=0.32\textwidth]{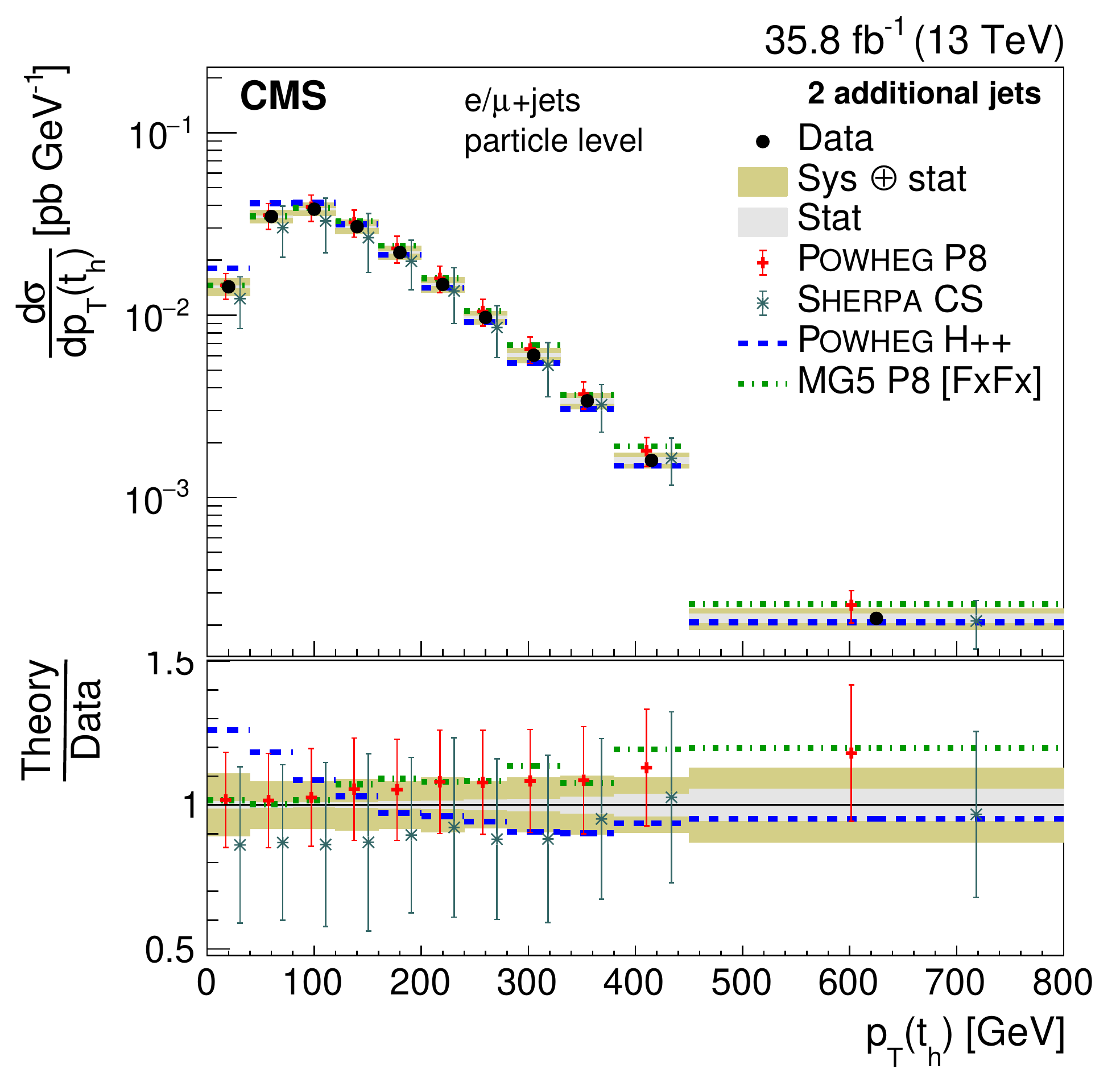}
\caption{Differential measurements as function of $\pt(\tq)$ at particle level in bins of jet multiplicities~\cite{lepjet13}. }
\label{fig_jetmulti}
\end{figure}

In Fig.~\ref{fig_jetmulti} a measurement of $\pt(\tq)$ in bins of multiplicities of additional jets is shown. Additional jets refers to jets with $\pt > 30$\,GeV and $|\eta| < 2.4$ that are not identified as one of the four jets in the \ttbar system. The softer measured $\pt(\tq)$ spectrum is most apparent if there is no additional jet. The spectra in events with one or more additional jets are better described by most of the simulations. 

\begin{figure}[h!!!]
\centering
\includegraphics[width=0.38\textwidth]{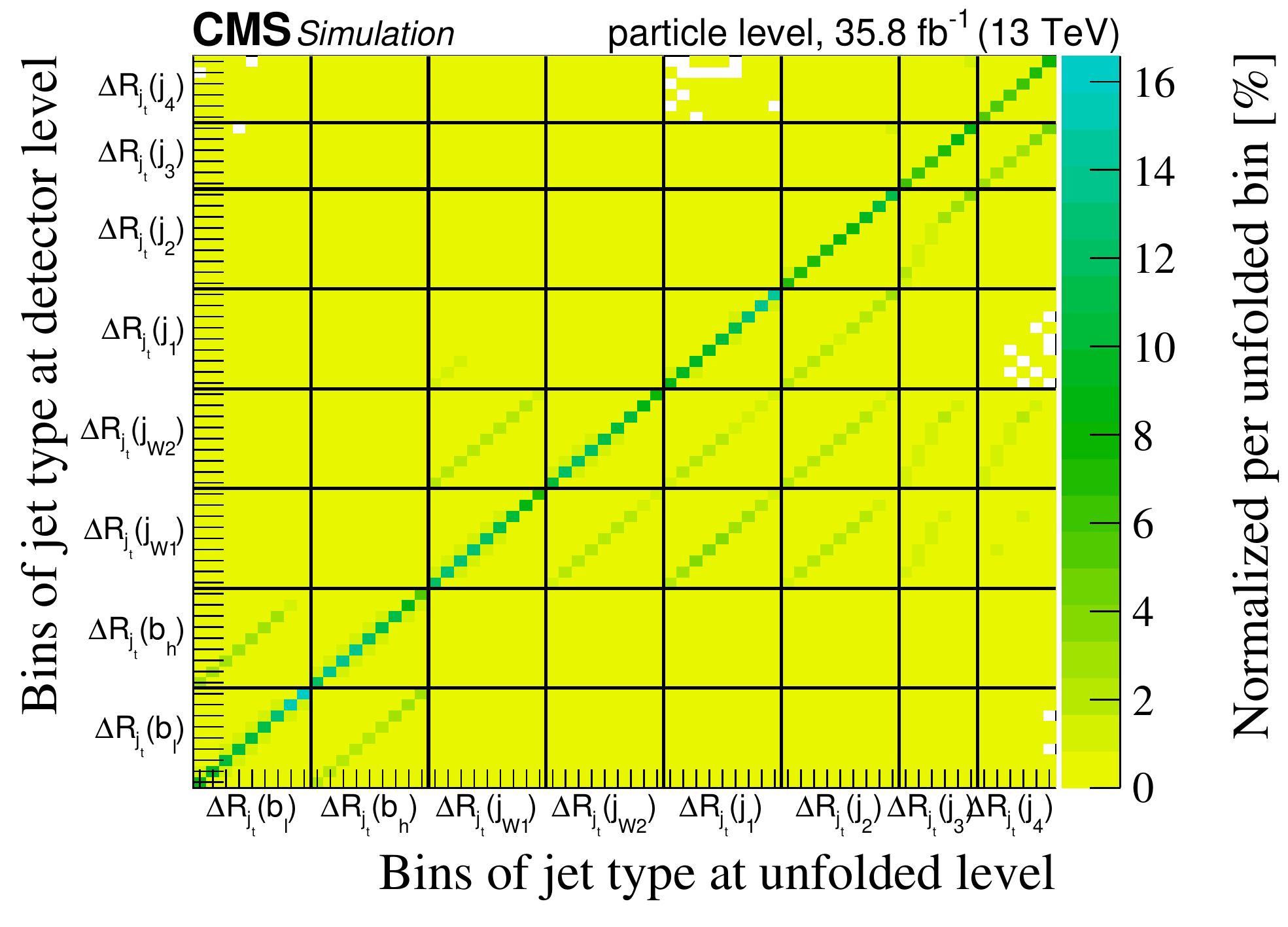}
\includegraphics[width=0.30\textwidth]{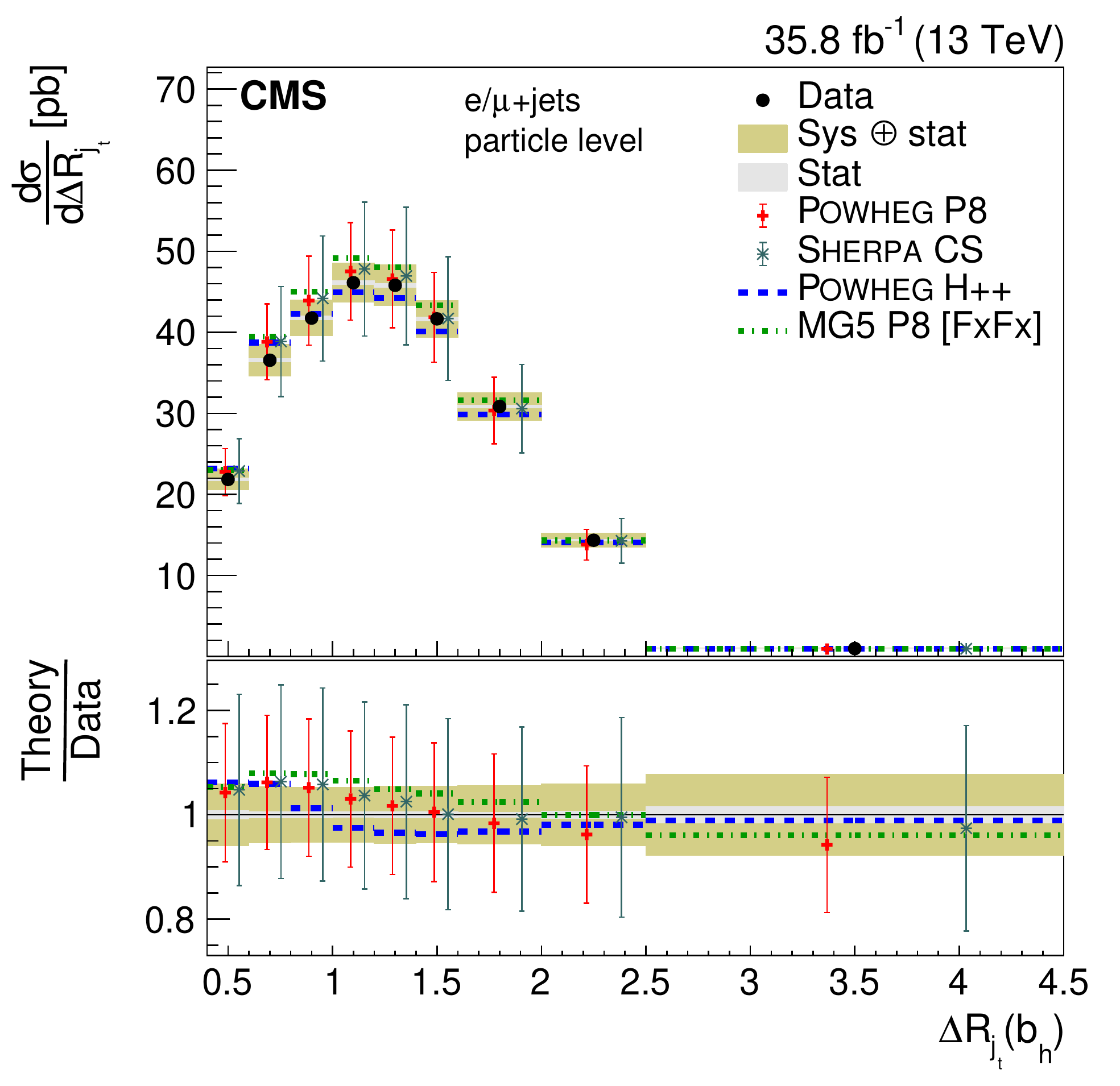}\\
\includegraphics[width=0.30\textwidth]{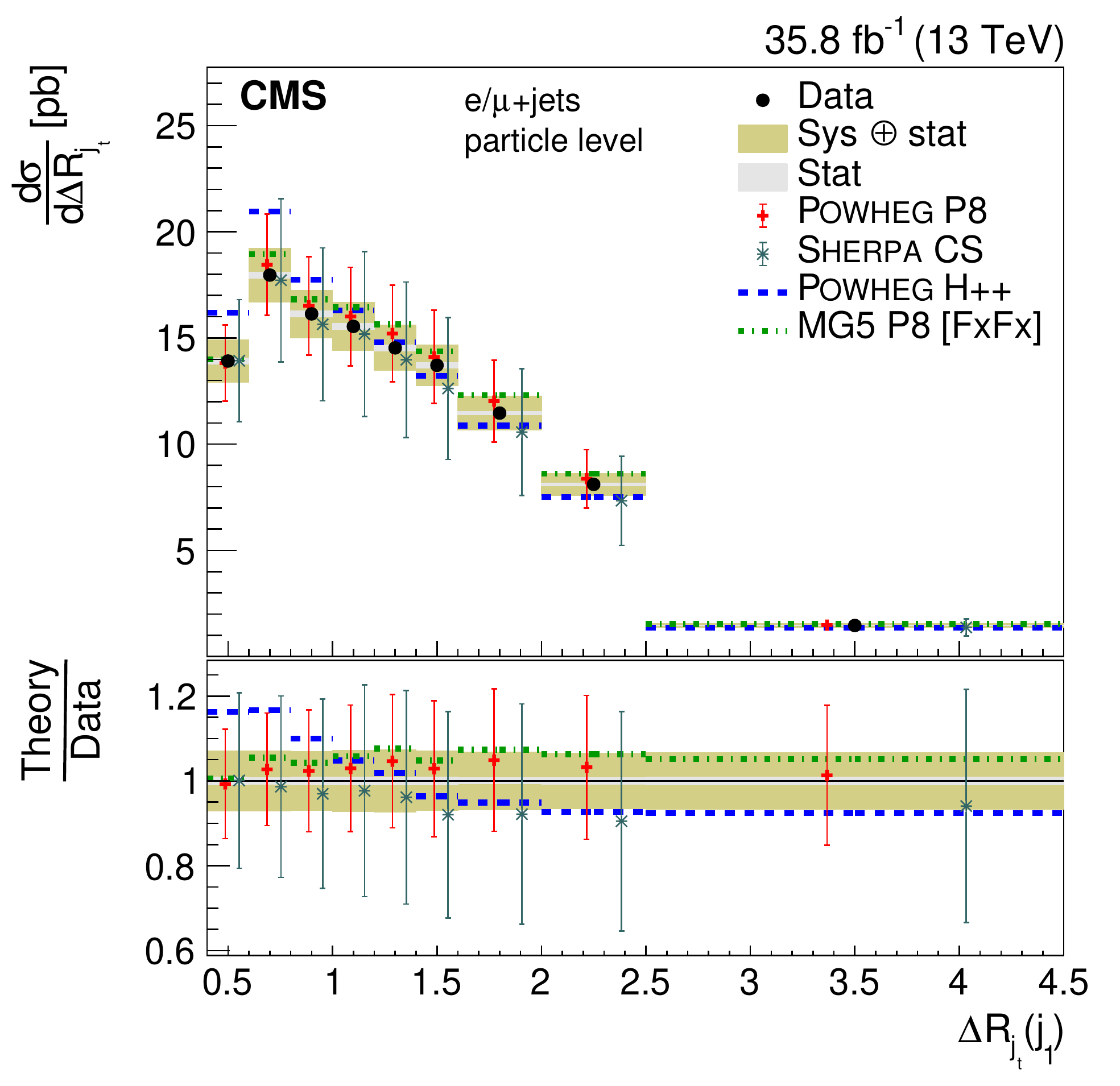}
\includegraphics[width=0.30\textwidth]{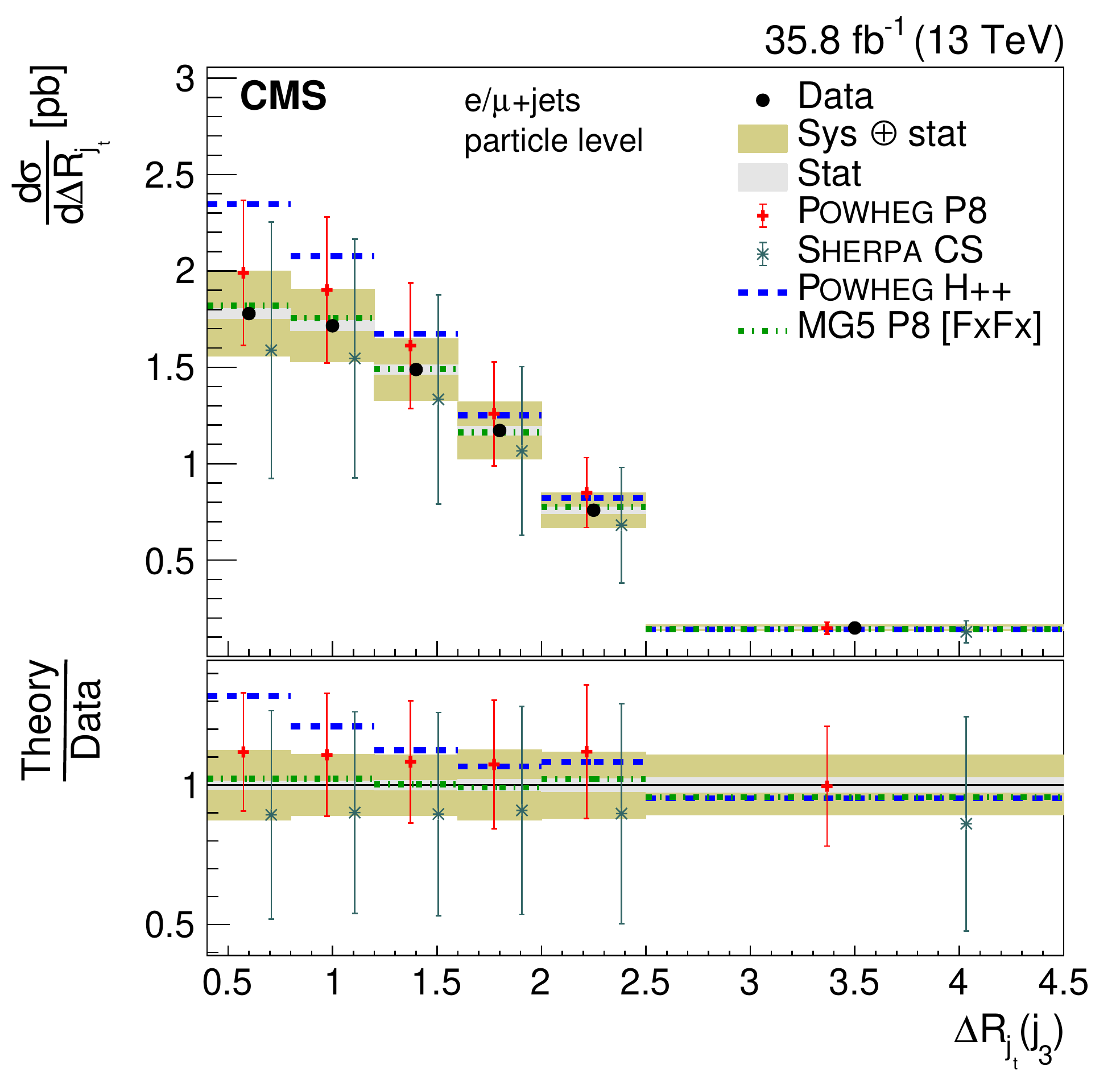}
\caption{Particle-level measurement as a function of \DRtopjets of the b jet stemming from the hadronically decaying top quark and as a function of \DRtopjets of the hardest and third hardest additional jets~\cite{lepjet13}.}
\label{fig_jetdr}
\end{figure}

Most of the previous figures show differences between the parton shower simulations of \PYTHIA (P8) and \HERWIG (H++), both in combination with the same next-to-leading-order matrix-element calculation of \POWHEG. To study further these differences and the effects of the parton shower modeling in general, kinematic properties of individual jets in \ttbar events are studied. This includes the kinematics of jets in the \ttbar system and of up to four additional jets. The unfolding of the detector-level spectra to the particle level does not only correct for resolution effects of the kinematic variable under consideration, but also applies corrections for the misidentification of the jets. A typical migration matrix is shown in Fig.~\ref{fig_jetdr} together with the measurement of the distance of jets from the closest jet in the \ttbar system \DRtopjets. This measurement shows an excess of jets close to jets in the \ttbar system in the \HERWIG simulation with the CMS tune and settings. This additional radiation reduces the $\pt(\tq)$ predicted by \HERWIG at particle level (Fig.~\ref{fig_pla}), while the $\pt(\tq)$ at parton level is not affected by the final-state radiation and describes the measurement quiet well (Fig.~\ref{fig_lepjet}).          

\end{document}